\documentclass[prl,twocolumn,showpacs,preprintnumbers,amsmath,amssymb]{revtex4-1}

\usepackage{graphicx}
\usepackage{dcolumn}
\usepackage{bm}

\begin{document}

\newcommand{\spinup}{\protect{$ \left|\uparrow \right\rangle$}}
\newcommand{\spindown}{\protect{$ \left|\downarrow \right\rangle$}}
\newcommand{\upeq}{\protect{\left | \uparrow \right\rangle}}
\newcommand{\downeq}{\protect{\left | \downarrow \right\rangle}}
\newcommand{\upxeq}{\protect{\left | \uparrow_x \right\rangle}}
\newcommand{\downxeq}{\protect{\left | \downarrow_x \right\rangle}}

\newcommand{\uu}{\protect{$ \left|\uparrow\uparrow\right\rangle$}}
\newcommand{\dd}{\protect{$ \left|\downarrow\downarrow \right\rangle$}}
\newcommand{\ud}{\protect{$ \left|\uparrow\downarrow \right\rangle$}}
\newcommand{\du}{\protect{$ \left|\downarrow\uparrow \right\rangle$}}
\newcommand{\uue}{\protect{\left|\uparrow\uparrow\right\rangle}}
\newcommand{\dde}{\protect{\left|\downarrow\downarrow \right\rangle}}
\newcommand{\ude}{\protect{\left|\uparrow\downarrow \right\rangle}}
\newcommand{\due}{\protect{\left|\downarrow\uparrow \right\rangle}}

\title{Spontaneous nucleation and dynamics of kink defects in zigzag arrays of trapped ions}

\author{S. Ejtemaee and P.~C. Haljan}
\affiliation{Department of Physics, Simon Fraser University,
Burnaby, BC, V5A 1S6, Canada}

\date{February 11, 2013}

\begin{abstract}
The spontaneous nucleation and dynamics of topological kink defects
have been studied in trapped arrays of 41--43 Yb ions. The number of
kinks formed as a function of quench rate across the linear-zigzag
transition is measured in the under-damped regime of the
inhomogeneous Kibble-Zurek theory. The experimental results agree
well with molecular dynamics simulations, which show how losses mask
the intrinsic nucleation rate. Simulations indicate that doubling
the ion number and optimization of laser cooling can help reduce the
effect of losses.  A range of kink dynamics is observed including
configural change, motion and lifetime, and behavioral sensitivity
to ion number.
\end{abstract}


\maketitle

The nucleation of topological defects during a symmetry-breaking
phase transition has many physical realizations. For continuous
phase transitions, the Kibble-Zurek mechanism (KZM) provides an
intuitive model of domain and defect formation~\cite{Kibble1976a,
Zurek1985a} and predicts a power-law scaling for the number of
defects formed as a function of transition quench
rate~\cite{Zurek1985a, Laguna1997a, Laguna1998a}. Several
experimental tests of the KZM have been done with different degrees
of success~\cite{KibbleReview, Chuang1991a, Bowick1994a,
Hendry1994a, Dodd1998a, Ruutu1996a, Bauerle1996a, Ducci1999a,
Maniv2003a, Monaco2006a, Monaco2009a, Sadler2006a, Scherer2007a,
Weiler2008a, Chen2011a, Chae2012a, Griffin2012a}, all with the goal
of examining universal behavior across diverse physical systems.

Following a recent proposal~\cite{delCampo2010a}, we report on the
nucleation and dynamics of topological defects in arrays of trapped
ions, following a quench of the linear-zigzag transition. Such
studies are well suited to this experimental system, since it is
simple enough to make direct comparison to theory, highly
controllable, and efficient in that the defects are imaged
\textit{in situ} and the trapped sample can be recycled repeatedly
across the transition.

\begin{figure}[t]
\centering
\includegraphics[width=\linewidth,clip]{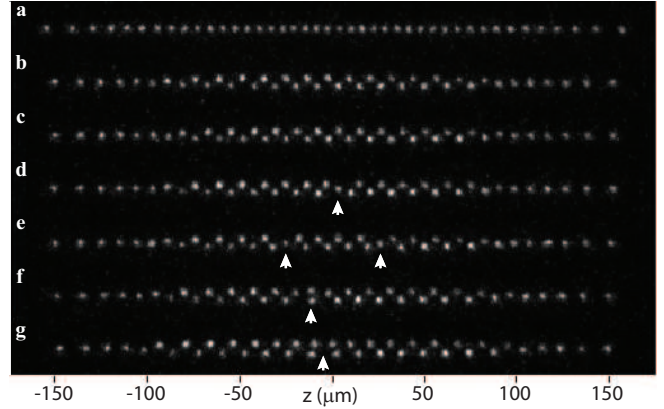}
\caption{Images of crystals of 42 $^{174}$Yb ions in various
configurations: (a) 1D linear crystal, (b) ``zig", (c) ``zag", (d)
zigzag with a single discrete kink (indicated by arrow), (e) two
discrete kinks (kink and anti-kink), and (f)--(g) single extended
kinks. The axial trap frequency for all images is
$\omega_z/2\pi\!=\!37.6$ kHz, while the weak transverse frequency,
$\omega_y/2\pi$, is (a) 658 kHz, (b)--(d) 414 kHz, (e) 380 kHz and
(f) 342 kHz. The corresponding quench end-points, $V_f$, are
(b)--(d) 4.8 V, (e) 5.3 V, and (f) 5.8 V. The imaging system looks
at 45$^{\circ}$ to the transverse $y$-axis where the zigzag
displacement occurs.} \label{fig:kinkimages}
\end{figure}

Laser-cooled ions held in a linear Paul trap~\cite{Raizen1992a} with
strong transverse confinement organize into a one-dimensional (1D)
linear crystal with an inhomogeneous axial density
[Fig.~\ref{fig:kinkimages}(a)]. If the transverse confinement is
relaxed slowly, the linear ion crystal undergoes a continuous,
structural phase transition to a 2D zigzag configuration, beginning
at the center of the crystal~\cite{Raizen1992a, Schiffer1993a,
Dubin1993a, Dubin1999a, Enzer2000a, Fishman2008a}. Two
broken-symmetry states are possible [Fig.~\ref{fig:kinkimages}(b) or
(c)]. Rapidly quenching the transverse confinement across the
transition can lead to crystal structures containing spontaneously
nucleated topological defects (kinks), which are formed by the
interface between domains of opposite symmetry
[Figs.~\ref{fig:kinkimages}(d)--(g)]~\cite{Landa2010a,delCampo2010a,FKModel}.
Del Campo\textit{ et al.}~\cite{delCampo2010a,Chiara2010a} have
applied the KZM formalism to the non-equilibrium dynamics of the
quench and calculated the number of kinks formed, which depends on
the quench rate and the damping rate ($\eta$) of the system,
controlled by Doppler laser cooling. In the under-damped regime,
where the system's response time is dominated by the inertial
dynamics of the soft transverse zigzag
mode~\cite{Laguna1998a,Fishman2008a}, the rate of defect formation
is independent of $\eta$ and given by a power law, $(2\delta_0/2\tau
_Q)^b$, where $2\tau_Q$ is the total quench time, and $2\delta_0$
defines the magnitude of the quench, which is linear in terms of
quadratic transverse trap frequency. The power-law scaling is
affected by the system inhomogeneity~\cite{IKZMOther,
delCampo2010a}, an effect which is largely unexplored
experimentally. Due to the inhomogeneous density, the propagation
speed of the transition front decreases from the crystal center to
the edges. This limits the size of the region where non-adiabatic
dynamics, and so nucleation, is possible, and for the under-damped
regime (${\eta}^3\!\ll\!\delta_0/\tau _Q$), gives a power-law
scaling of $b=4/3$ (IKZM regime)~\cite{delCampo2010a}. Following
calculations in other
contexts~\cite{Saito2007a,Dziarmaga2008a,Monaco2009a}, it has been
pointed out that for small numbers of ions or slow quenches, where
the size of the nucleation region is smaller than the correlation
length, the scaling doubles to $b=8/3$ (DIKZM
regime)~\cite{Pyka2012a}.

Simultaneous experiments to ours have recently reported kink
nucleation in ion arrays due to a temperature quench of a plasma
cloud~\cite{Mielenz2012a}, and a measurement of the DIKZM power
law~\cite{Pyka2012a}. Compared to the latter results, our
measurements of nucleation are also due to a quench across the
structural transition but use a different experimental approach,
including more ions and a shallower quench into the zigzag phase so
that kinks retain their localized, discrete form. Moreover, we
explore post-nucleation dynamics of discrete kinks including
transformation of shape, motion and lifetime in the crystal, and
sensitivity of behavior to unit change in ion number.


The experimental set-up is similar to that previously
reported~\cite{Ejtemaee2010a, Liang2011a}. We use a linear Paul
trap, operating at a radio frequency (rf) of
$\Omega_{r\!f}/2\pi\!=\!12$ MHz, and composed of four rods with an
ion-to-rod spacing of 0.7 mm and two end-caps separated by 2.5 mm,
to form a nominally harmonic confining potential for ions.
Photo-ionization is used to load a pure crystal of 42--43 $^{174}$Yb
ions~\cite{Balzer2006, Liang2011a}, which arrange as a linear string
in a trap with an axial secular frequency of
$\omega_z/2\pi\!=\!37.6$ kHz and transverse frequencies of
$\omega_x/2\pi\!=\!679$ kHz and $\omega_y/2\pi\!=\!658$ kHz
[Fig.~\ref{fig:kinkimages}(a)]. A single laser beam at 369.5 nm
(with axial FWHM of 280 $\mu$m) is used for fluorescence imaging of
the ions and for Doppler cooling along all principal axes of the
trap. The 369.5-nm laser is linearly polarized at $6^\circ$ with
respect to a 5.9-G magnetic field. A relatively large laser detuning
of $-33$ MHz (1.5 linewidths) is used to suppress the effects of
micromotion on fluorescence and cooling~\cite{Berkeland98}, arising
from a non-zero component of the trap rf electric field along the
axial direction (up to 2500 V/m at the edges of a 43-ion crystal).
With an effective saturation parameter of $s_{0}\!=\!2.7$, the
estimated cooling rate is uniform to a factor of 2 or better across
the crystal with variation primarily at the edges, and the
calculated cooling time, $\eta^{-1}$, at the center of the crystal
is 200--400 $\mu$s~\cite{Itano1982a}. The cooling rate places the
experiment deeply in the under-damped regime for quench rates
considered.

To quench the system through the linear-zigzag transition, we apply
a transverse dc quadrupolar potential using the trap rods.  The
secular frequency along the transverse $y$-axis, obtained from
single-ion calibrations, decreases with the applied dc voltage,
$V(\geq0)$, according to $\omega_y(t)=\omega_{y0}\sqrt{1-V/V_{0}}$
where $\omega_{y0}/2\pi\!=\!658(1)$ kHz and $V_{0}$=7.95(2) V with
errors including drift. The $x$-axis frequency increases in an
equivalent way. The quench waveform is linear ($V(t)=V_f\cdot
t/(2\tau_Q)$) with a quench end-point of $V_f\!=\!4.2-5.8$ V, a ramp
time of $2\tau_Q\!=\!10-70$ $\mu$s, and a maximum distortion of 50
mV$\!_{pk}$ after filtering. The quench also incorporates small
adjustments to the end-caps and rods to maintain a constant axial
secular frequency and to maintain the center of the crystal at the
rf null in all dimensions. The linear form of our quench in terms of
$\omega_y^2$ is as assumed in the KZM theory~\cite{delCampo2010a},
with the one modification that we use an asymmetric ramp across the
critical point. For example, in a 42-ion string with
$\omega_z/2\pi\!=\!37.6$ kHz, where the critical transverse
frequency at the center of the crystal is
$\omega_y^c(0)/2\pi\!=\!619(3)$ kHz (or equivalently, 0.92(6)
V)~\cite{Shift}, our quench starts at 658 kHz ($V_i\!=\!0.0$ V),
which ensures a sufficiently linear configuration~\cite{SoftMode},
and continues further past the critical point, for example to 414
kHz ($V_f\!=\!4.8$ V) to involve about half of the crystal in the
zigzag phase.

Each experimental run begins with 53 ms of cooling in the linear
configuration, followed by the quench (with cooling always active).
Images of the initial linear and subsequent zigzag configurations
are recorded with an exposure time of 3 ms. A 200-$\mu$s ramp
recycles the crystal to the linear structure. Data is acquired in
batches of 100--700 runs over 3 minutes. Our data sets contain 80\%
of runs approximately equally divided between 42 and 43 ions with 41
or fewer ions in the remainder. Occasionally, the fluorescence of an
ion is interrupted by internal-state pumping or molecule formation
through background-gas collisions~\cite{Liang2011a}. Runs with dark
ions within the zigzag region of the crystal ($\leq$ 14\% of runs,
typical) are excluded. Image analysis is used to identify locations
of any kinks in the zigzag structure. For configurations without
kinks, temporal statistics of the two symmetry-broken phases are
nominally random with a bias of $\lesssim$1\% and a normalized
binary autocorrelation on adjacent runs of $\lesssim$1\%.

\begin{figure}[t]
\centering
\includegraphics[width=\linewidth,clip]{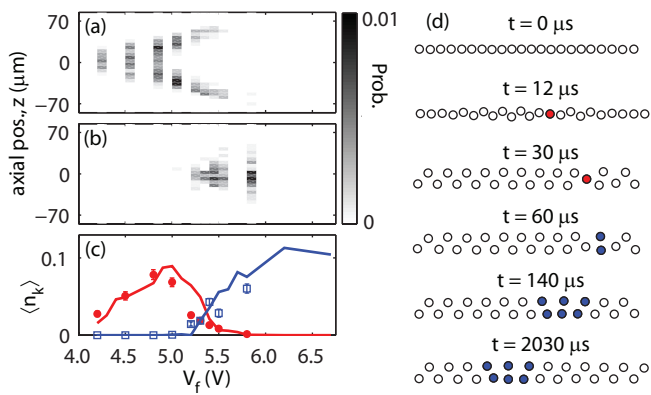}
\caption{(color online) Observed axial position distribution of (a)
discrete and (b) extended kinks versus quench end-point. All data
taken at a quench rate of 0.19 V/$\mu$s. (c) Average number per run
of discrete kinks (red circles) and extended kinks (blue squares)
for distributions in (a) and (b). Data points, for 41--43 ions,
include 1000--2200 runs. Errors calculated assuming binomial
statistics. Lines are a simulation for 42 ions with no free
parameters, and shown 2 ms after the quench, 2/3 of the camera
exposure time. Statistical uncertainties in simulation comparable to
experiment ones shown. (d) Simulation frames showing evolution after
nucleation of a discrete kink (red) into an extended kink (blue) at
$V_f = 5.8$ V ($\omega_y/2\pi$=342 kHz). Middle 144 $\mu$m of a
42-ion crystal shown. }
 \label{fig:kinktype}
\end{figure}

In the first experiment, we explore the formation of two types of
kink~\cite{Landa2010a}: discrete kinks, which are localized to about
3 ion-sites [Figs.~\ref{fig:kinkimages}(d),(e)], and extended ones,
which involve more sites and occur in crystals brought far into the
zigzag region [Figs.~\ref{fig:kinkimages}(f),(g)].
Figures~\ref{fig:kinktype}(a)--(c) show the observed axial
distribution of both types and their total numbers as a function of
the quench end-point, $V_f$. The data, for 41-43 ions, is taken at a
constant ramp rate of 0.19 V/$\mu$s so that the nucleation dynamics
near the critical point is expected to be same. Any differences are
due to post-nucleation dynamics during the remainder of the ramp
through to detection. To understand the outcome of the experiment,
we have performed molecular dynamics simulations with no free
parameters: The ion trap is treated as an ideal harmonic potential
limited to two dimensions. The small axial anharmonicity in the
experiment is modeled in an effective way by shifting the axial
frequency from 37.6 to 38.2 kHz~\cite{Shift}. The effect of the
Doppler cooling laser is modeled explicitly~\cite{Casdorff1988a} and
includes the steady-state fluorescence behavior of $^{174}$Yb$^+$
and all laser calibrations excluding beam
profile~\cite{Ejtemaee2010a}. Simulations for 42 ions reveal similar
distributions to Figs.~\ref{fig:kinktype}(a) and (b). There is also
a good qualitative match in the numbers of discrete and extended
kinks between simulation and experiment [see
Fig.~\ref{fig:kinktype}(c)]. From the simulations, we find that
discrete kinks are nucleated in the central 5-10 sites of the
crystal. Then, there is an outward motion and loss of the discrete
kinks, propelled by the gradient in zigzag amplitude and axial
density, which affect the underlying potential for the kinks. As the
quench end-point increases towards 5.0 V, the number of discrete
kinks and the width of their distribution increase since more of the
dispersing kinks are trapped in a deeper and wider zigzag region.
Above 5.0 V, the number of discrete kinks collapses when, as seen in
the simulation, those at the center of the crystal distort to
extended form. Extended kinks can also migrate to the center
[Fig.~\ref{fig:kinktype}(d)], suggesting a change in the underlying
potential for the kinks and leading to the well-centered
distribution observed in the experiment
[Fig.~\ref{fig:kinktype}(b)]. Due to reduced losses, the number of
extended kinks in the simulation saturates at a value 1.4 times
higher than the peak in discrete number. The experiment data ends at
5.8 V, well before the $\sim$7.1-V onset of another structural
transition to a different 2D configuration. As this transition is
approached, we see an enhancement in (otherwise negligible)
background kink nucleation, due to energy transfer events such as
background-gas collisions.

\begin{figure}[t]
\centering
\includegraphics[width=\linewidth,clip]{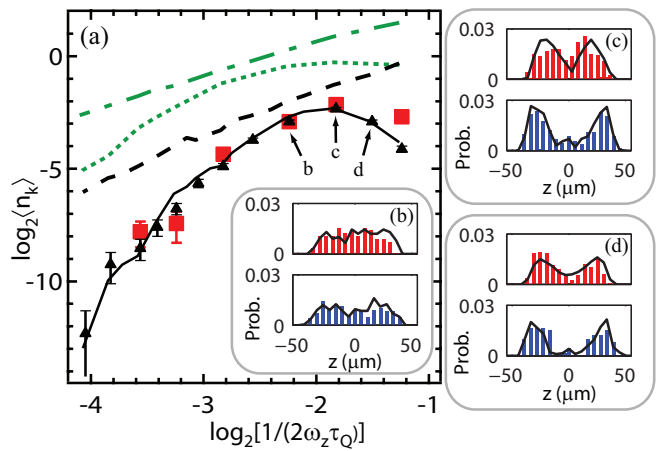}
\caption{(color online) (a)Average number of kinks observed per run
vs. scaled quench rate for standard trap with $\omega_z/2\pi = 37.6$
kHz (triangles), and for one with secular frequencies reduced by 2
(red squares) and quench changed accordingly. Data, for 41--43 ions,
include 1000--5000 runs. Error bars are calculated binomial,
approximately consistent with fluctuations. Solid line is a
molecular dynamics simulation with parameters including
$\omega_z/2\pi$ = 38.2 kHz (see text), and shown 2 ms after the
quench. Statistical uncertainties of simulation less than or equal
to the experiment. Dashed black line is the same simulation with
kinks counted shortly after nucleation ($\geq$7 $\mu$s). To suppress
thermal fluctuations, the kink search is limited to the zigzag
region with minimum amplitude of 0.54 $\mu$m. Dotted and dash-dotted
lines (green) show kink number 2 ms after quench and shortly after
nucleation, respectively, for a simulation of 84 ions with optimized
Doppler cooling (detuning $-$11.3 MHz, $s_0$ = 2.0) and with
$\omega_z/2\pi$ = 20.6 kHz, chosen to match the transverse critical
frequency for 42 ions at crystal center. Simulation plot scaled
using $\omega_z/2\pi$ = 38.2 kHz for direct comparison to 42 ions.
(b)--(d) Observed position distributions of kinks for 42 ions (red,
upper) and 43 (blue, lower) in standard trap at quench time
$2\tau_q$ of (b) 20, (c) 15 and (d) 12 $\mu$s with simulations
(lines) for comparison.} \label{fig:ramp_plot}
\end{figure}

Next, we proceed to study kink nucleation as a function of quench
rate. With an eye towards subsequent dynamics studies, we work with
discrete kinks and operate at $V_f\!=\!4.8$ V, where their maximum
number is detected [Fig.~\ref{fig:kinktype}(c)].
Figure~\ref{fig:ramp_plot}(a) shows a log-log plot of the average
number of kinks detected per run (black triangles) as a function of
scaled rate, $1/(\omega_z\cdot2\tau _Q$). Data for 41--43 ions is
combined. As the quench rate increases from $1/(\omega_z\cdot70
\mu$s), the number of kinks grows, peaks at $1/(\omega_z\cdot15$
$\mu$s) and begins to drop. Figure~\ref{fig:ramp_plot}(a) also
includes results of molecular dynamics simulations, performed as
described above. The first simulation (solid black line) shows the
number of kinks 2 ms after the quench (2/3 of the exposure time in
the experiment), by which time the kink number has stabilized. The
simulation matches the observed number of kinks extremely well, and
reproduces features in the observed kink distributions, related to
rapid, post-nucleation dynamics [Fig.~\ref{fig:ramp_plot}(b)--(d)].
To understand the effect of kink losses before detection in the
experiment, we use the simulation to assess the ``early-counted"
number of kinks shortly after nucleation (dashed black line). At
fast quenches, there is enhanced loss due to kinetic energy from
nonadiabatic excitation of vibrational modes, and before the crystal
is re-cooled. Kink pair annihilation also plays a minor role. At
slow quenches, we again see a loss of kinks, which worsens with
decreasing quench rate. In this case, the zigzag region expands too
slowly to trap the kinks, and they escape to the zigzag edges during
the quench.

Although kink losses in the experiment mask the intrinsic nucleation
rate -- see, however, Ref.~\cite{Pyka2012a} -- we comment briefly on
the power law scaling for the simulation and experiment.
Figure~\ref{fig:ramp_plot}(a) covers about three octaves in quench
rate separated by $\log_2 (1/2\omega_z\tau_Q)\!=\!-3.1$ and $-$2.1.
The KZ theory~\cite{delCampo2010a,Pyka2012a} suggests that our data
covers two regimes of nucleation: For the slow-quench range, we
expect a DIKZM power law of $8/3\!=\!2.67$, which agrees with the
fit value of 2.5(2) for the early-counted simulation (with
significant systematic error associated with rapid losses).  For the
fast-quench range, where the nucleation region is large enough that
nucleation of two or more kinks becomes significant in the
simulation, one expects to cross over towards the IKZM power law of
$4/3\!=\!1.33$. The early-counted simulation gives a fit value of
1.7(1). We note that all cases of two kinks observed in the
experiment [Fig.~\ref{fig:kinkimages}(d)], albeit rare, occur in or
adjacent to the fast-quench region. Finally, in the middle octave of
quench rate, where the kink losses are least, a fit to the
\textit{experimental} data gives a power law of 3.3(2), compared to
the early-counted simulation value of 2.0(1).

For comparison, we also measure kink nucleation in a trap with all
secular frequencies reduced by a factor of 2 (red squares in
Fig.~\ref{fig:ramp_plot}(a)). Overall, the data match results at
higher trap frequency, which shows the weak sensitivity of
nucleation and losses to effective cooling rate, $\eta/\omega_z$,
over such a variation, except perhaps for losses at the fastest
quenches~\cite{Insensitivity}. Of several possibilities to reduce
the losses in our experiment, we note that an increase in ion number
can help, particularly for slow quenches by reducing the crystal
inhomogenieties that transport discrete kinks outward. As shown in
Fig.~\ref{fig:ramp_plot}(a), a doubling of ion number (and using
twice better cooling) already leads to significant reduction in
losses over the same quench range, albeit in a scaling regime closer
to IKZM.

\begin{figure}[t]
\centering
\includegraphics[width=\linewidth,clip]{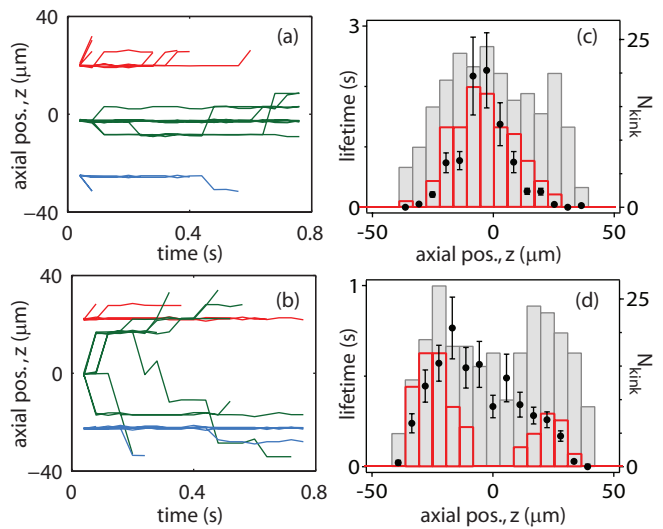}
\caption{(color online) (a)--(b) Axial trajectories of discrete
kinks in zigzag crystals ($V_f$ = 4.8 V) with (a) 42 and (b) 43 ions
for three different initial positions: center and four sites away on
both sides. The number of trajectories for each case is 11 with
fewer for $z<0$. (c)--(d) Exponential lifetime of kinks (filled
circles) vs initial position in crystals of (c) 42 and (d) 43 ions.
Lifetimes and their uncertainties obtained from a maximum likelihood
estimate including binning due to camera frame rate. Also shown are
axial position distributions at $t=0$ ms (filled gray bars) and
$t=400$ ms (open red bars).}
 \label{fig:dynamics}
 \end{figure}

We conclude with an investigation of the dynamics of single discrete
kinks, and measure their axial motion in the crystal and their
lifetime. The experiments are performed at $V_f\!=\!4.8$ V and
$2\tau_Q=20$ $\mu$s, for which the initial distribution of the kinks
is relatively uniform [Fig.~\ref{fig:ramp_plot}(b)].
Figures~\ref{fig:dynamics}(a) and (b) show the trajectories of kinks
in crystals of 42 and 43 ions for three initial axial positions. The
kinks move between sites corresponding to local potential wells
separated by the axial lattice spacing (5.6 $\mu$m)~\cite{FKModel}.
The 40-ms frame rate of our camera does not permit us to resolve
fast dynamics, including the motion between sites. Nevertheless, it
can be seen that, for 42 ions, most of the kinks that begin at the
center of the crystal either stay there or move by one site. In
comparison, the central kinks in a string of 43 ions migrate to the
edges faster. For both cases, kinks closer to the edges tend to move
outwards, as expected.

In Figs.~\ref{fig:dynamics}(c) and (d), we show the exponential
lifetime of kinks as a function of their initial position as well as
the distributions of kink positions at $t=0$ s and $t=0.4$ s. In
Fig.~\ref{fig:dynamics}(c), we can see that the lifetime of kinks
for 42 ions is highest near the middle of the crystal (with the
$\sim$2-s value likely limited by background-gas collisions), and
decreases roughly symmetrically towards the edges. Losses due to the
position-dependent lifetime largely account for the observed time
evolution of the position distribution. In
Fig.~\ref{fig:dynamics}(d), for 43 ions, the 0.3--0.5 s lifetime of
central kinks contrasts with their complete absence in the position
distribution after 0.4 s, and points to their migration away from
the deeper zigzag region at the center of the
crystal~\cite{LifetimeAsymm}. Simulations show that this is
associated with new dynamics in which the kink hops two sites (as
observed in Fig.~\ref{fig:dynamics}(b) center) via a transient
extended-like form. The difference between 42 and 43 ions is
understandable, considering that the experiment is performed at
$V_f=4.8$ V, near to the sharp onset of the bimodal distribution in
Fig.~\ref{fig:kinktype}(a). Since the zigzag phase begins 0.3 V
lower for 43 ions, the onset of the bimodal distribution will also
occur at a lower voltage.

To conclude, we have measured the number of kinks formed in ion
crystals as a function of quench rate in the under-damped regime of
the KZM, and have shown how losses mask the intrinsic nucleation
rate. Although simultaneous results based on a deeper quench and
extended kinks show lower loss~\cite{Pyka2012a}, our method of
nucleation assessment should be advantageous for larger ion numbers
with multiple kinks, since the localized character of the discrete
kinks will help reduce interaction-dependent losses. We have also
studied the post-nucleation dynamics of kinks, and have observed
long lifetimes, which may allow future explorations of internal kink
dynamics~\cite{Landa2010a,FKModel}. A natural future direction is
the study of linear-zigzag dynamics in the quantum
regime~\cite{Retzker2008a}.

\begin{acknowledgments}
The authors thank Malcolm Kennett and Jeff McGuirk for helpful
comments on the manuscript. This work is supported by NSERC and the
CFI LOF Program and has been enabled by computing resources provided
by WestGrid and Compute/Calcul Canada.
\end{acknowledgments}


\begin{thebibliography}{48}
\bibitem [{Shi()}]{Shift}%
  \BibitemOpen
  \href@noop {} {}\bibinfo {note} {The observed value of $\omega_y^c(0)/2\pi$
  is slightly higher than expected (1 percent), roughly consistent with
  measured axial anharmonicity vs. ion number, and is accounted for in
  simulations by a shift in $\omega_z$.}\BibitemShut {Stop}%
\bibitem [{Sof()}]{SoftMode}%
  \BibitemOpen
  \href@noop {} {}\bibinfo {note} {The soft--mode frequency at quench start is
  also larger than the KZM freeze-out value. Simulations of quenches starting
  at higher transverse frequency (883 kHz) give quantitatively similar kink
  numbers to those observed.}\BibitemShut {Stop}%
\bibitem [{Ins()}]{Insensitivity}%
  \BibitemOpen
  \href@noop {} {}\bibinfo {note} {The overall weak sensitivity shown also
  supports ignoring small (spatial) variations in cooling, corroborated by good
  agreement with simulations neglecting them. Note the cooling beam is also
  larger for this experiment.}\BibitemShut {Stop}%
\bibitem [{Lif()}]{LifetimeAsymm}%
  \BibitemOpen
  \href@noop {} {}\bibinfo {note} {Asymmetries in the lifetime distributions
  are likely due to experimental imperfections. Longer lifetime of outer kinks
  for 43 vs. 42 ions and changes in dynamics with ion number and density are
  interesting and will be explored in a future work.}\BibitemShut {Stop}%
\end{thebibliography}

%



\end{document}